\documentclass{PoS}

\title{Reconciling radio relic observations and simulations: The NVSS sample}

\ShortTitle{The NVSS radio relic sample}

\author{\speaker{Jakob Gelszinnis}\\
        Th\"uringer Landessternwarte Tautenburg (TLS), Sternwarte 5, D-07778, Tautenburg, Germany\\
        E-mail: \email{jakobg@tls-tautenburg.de}}

\author{Matthias Hoeft\\
       Th\"uringer Landessternwarte Tautenburg (TLS), Sternwarte 5, D-07778, Tautenburg, Germany\\
       E-mail: \email{hoeft@tls-tautenburg.de}}   
       
\author{Sebasti\'an E. Nuza\\
       Leibniz-Institut f\"ur Astrophysik Potsdam (AIP), An der Sternwarte 16, D-14482 Potsdam, Germany \\
       E-mail: \email{snuza@aip.de}}

\abstract{The diffusive shock acceleration scenario is usually invoked to explain radio relics, although the detailed driving mechanism is still a matter of debate.
          Our aim is to constrain models for the origin of radio relics by comparing observed relic samples with simulated ones. 
         Here  we present a framework to homogeneously extract the whole sample of known radio relics from NVSS  so that it can be used for comparison with cosmological simulations.
         In this way, we can better handle intrinsic biases in the analysis of the radio relic population.   
         In addition, we show some properties of the resulting NVSS sample relics such as the correlation between relic shape and orientation with respect to the cluster.
         Also, we briefly discuss the typical relic surface brightness and its relation to projected cluster distance and relic angular sizes.
           }

\FullConference{EXTRA-RADSUR2015 (*)\\
		20--23 October 2015\\
		Bologna, Italy

                \bigskip
                \hrule
                \bigskip

                \textnormal{(*) This conference has been organized
                  with the support of the Ministry of Foreign Affairs
                  and International Cooperation, Directorate General
                  for the Country Promotion (Bilateral Grant Agreement
                  ZA14GR02 - Mapping the Universe on the Pathway to
                  SKA)}
}

\begin{document}

\section{Motivation}
There exists a class of synchrotron emitting structures on cluster scales for which no galactic 
counterpart can be identified known as radio relics. 
The fact that they are predominantly found in merging systems leads to the belief that they 
are produced when electrons become accelerated at, or shortly behind, the resulting shock fronts. 
Nowadays, around 50 clusters hosting those objects are known, with the prospect of hundreds 
of new identifications within the next decade radio and cluster surveys \cite{Nuza+2012}.

It is still an ongoing matter of discussion if, or to which extent, diffusive shock acceleration (DSA) 
is the driving mechanism behind this phenomenon. Puzzling mismatches between the shock measurements in 
X-ray and radio observations, as well as the existence of bent spectra 
and unexpected electron-to-proton ratios in some relics, lead to the question whether 
additional physics is needed to explain these observations. 
Because synchrotron emission is well understood, it is rather a question about the origin 
of the underlying population of highly-relativistic electrons.
Both efficient accelerated mechanisms at weak shocks, like shock drift acceleration, 
and pre-existent non-thermal electron populations 
could be needed to explain the observed discrepancies.

Here we describe some basic properties of the whole relic sample within NVSS \cite{NVSS1998}. This sample has been homogeneously constructed to further compare with a set of simulated DSA relics in an attempt to constrain their origin.
While this approach is quite simplistic, it is clear that it is neither restricted to a specific radio survey, nor frequency, nor theoretical model of radio relic formation.
A combined analysis of the NVSS and MUSIC-2 cluster simulation \cite{MUSIC:I}  will be presented in a forthcoming paper (Nuza et al.\,in prep.).

\section{Measuring radio relics in NVSS}

In order to compare with simulations it is necessary to produce a statistically 
significant observed sample of relics with relatively low errors 
and known detection and systematic biases.

\begin{figure}[htp]
  \includegraphics[width=0.64\linewidth]{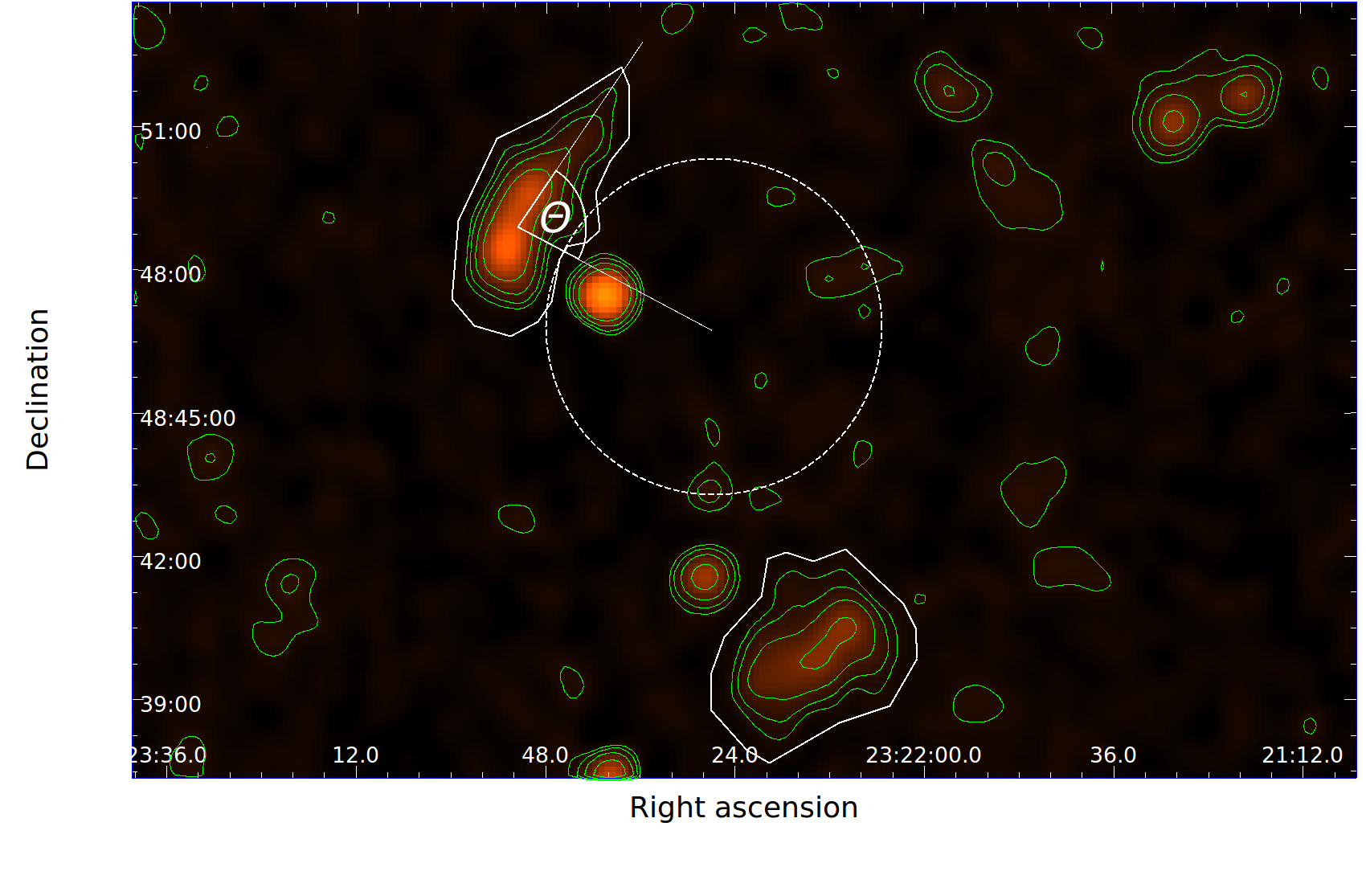}
  \includegraphics[width=0.37\textwidth]{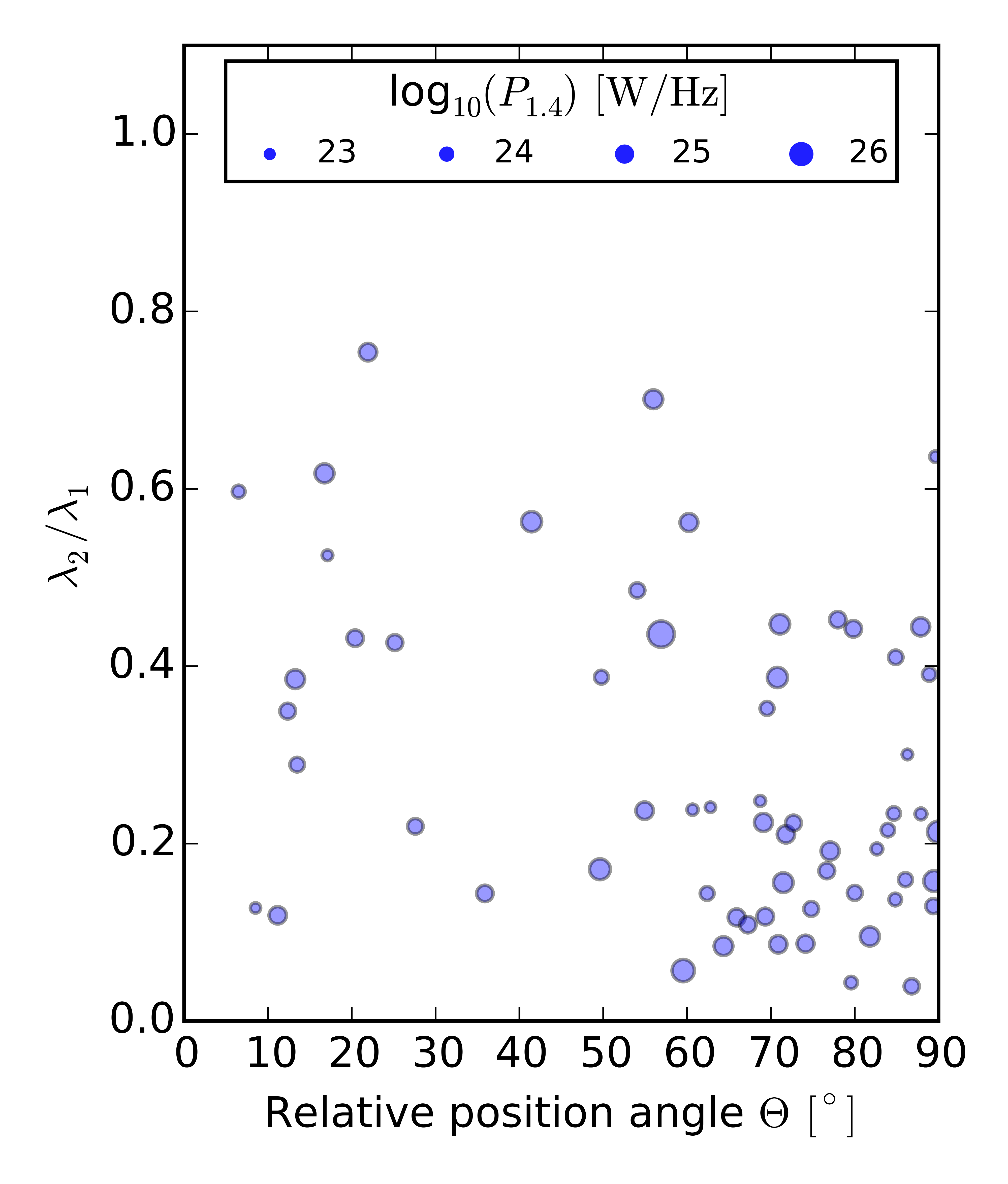}
  \caption{NVSS image of the double radio relic in PSZ1 G108.18-11.53 (left-hand panel, see \cite{DeGasperin+2015}). 
  Radio brightness is also shown as contours spaced by 
  $2^n\times\sigma_\mathrm{rms,NVSS}$ with $n=[1,2,3,4]$. 
  The relic search area is shown as white polygon. 
  A projected physical distance of 1\,Mpc to the cluster center is denoted by the dashed circle. 
  The flux-weighted axis joining the relic and cluster centers, the relic major axis, 
  and the angle between them are depicted. Relic shape  statistics versus the relic-cluster 
  alignment angle $\Theta$ is also shown (right-hand panel).}
\label{Fig:NVSSscheme}
\end{figure}

With many exciting new surveys on the horizon, NVSS is still the best fulfilling our criteria as 
it combines a large survey area with a resolution and frequency favourable for relic detection.  
The majority of known radio relics are identified within NVSS when applying a detection 
threshold of two times the survey noise, i.e. $0.9\,\mathrm{mJy/beam}$ at 1.4 GHz observed frequency
(see e.g., the left-hand panel of Fig.\,\ref{Fig:NVSSscheme}). 
We recovered relics from regions specified by the location indicated in the literature.
Detection and measurement of intrinsic properties were based on images with subtracted off 
compact sources identified through published high-resolution observations.

We assume NVSS, as well as any suitable mock observations, to show the same characteristics and design an automatic source extraction scheme valid for both types of samples.
Relics were extracted using their $2\sigma_\mathrm{rms}$ contours to measure their largest angular scale (LAS), position, shape, area, and luminosity.
For these measurements we extensively relied on the already widespread use of image moments \cite{Stobie1980}.
Relics with a total flux below four times the relic detection threshold, 
i.e. $3.6\,\mathrm{mJy/beam}$, were excluded because their shape is significantly affected by noise.

Our final sample contains 70 relic regions in 38 clusters above $-40^\circ$ latitude,
including several relic fragments otherwise identified as a single object.

\section{Relic geometries}
From here on, we focus on the radio \textit{gischt} subpopulation of relics which 
are primarily thought to trace cluster merger shocks (e.g. 
we exclude radio phoenixes, thought to stem from fossil radio plasma).

We characterize the relic shape, $s=\lambda_2/\lambda_1$, as the ratio of eigenvalues 
$\lambda_i$ of the image moment covariance matrix ($s=0,1$ for a line and a circle, respectively.) 
The right-hand panel of Fig.\,\ref{Fig:NVSSscheme} shows the shape parameter $s$ and the 
relic orientation angle. The Spearman's rank statistics yields a correlation coefficient 
of $\rho=-0.34$  and $p=6.4\times 10^{-3}$ for the null hypothesis, supporting 
the idea of an anti-correlation between those two variables.
This correlation was not described before and it suggests that 
relic shapes may be affected by their projection onto the cluster.

Also, we confirm the already reported strong tendency of relics to be orthogonally aligned 
with respect to the line joining the relic and cluster centers, 
i.e  large $\Theta$ values (see \cite{vanWeeren+2011}, their Fig.\,22). 
This is what one would expect if relics would trace shock fronts traveling outwards 
in the direction of the merger axis. We find that there are 12 relics with $\Theta<30^\circ$; 
9 with $30^\circ<\Theta<60^\circ$, and 41 with $\Theta>60^\circ$.

\section{Where does the flux reside?}
 
  \begin{figure}
    \includegraphics[width=0.49\linewidth, height = 5cm]{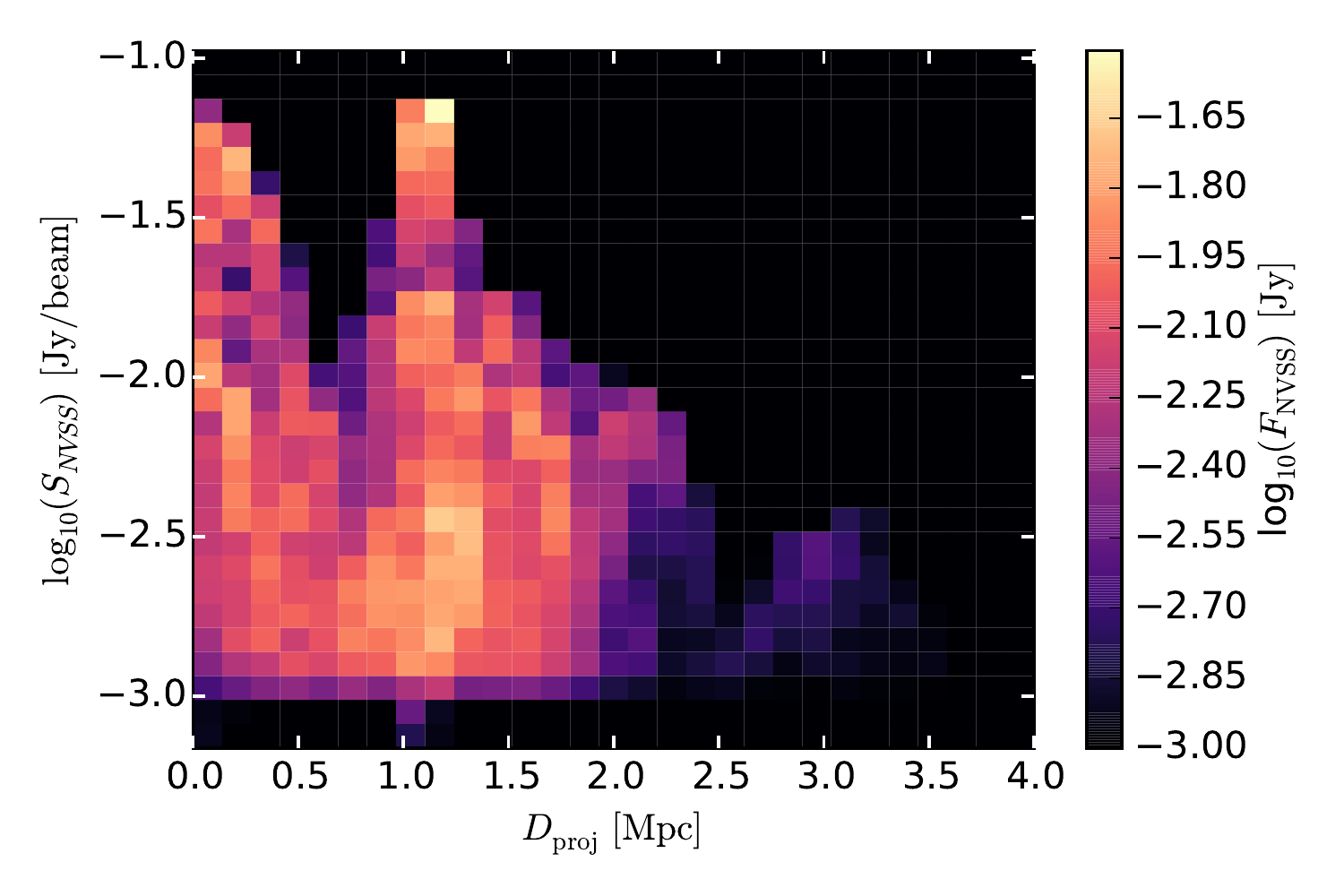}    
    \includegraphics[width=0.49\linewidth, height = 5cm]{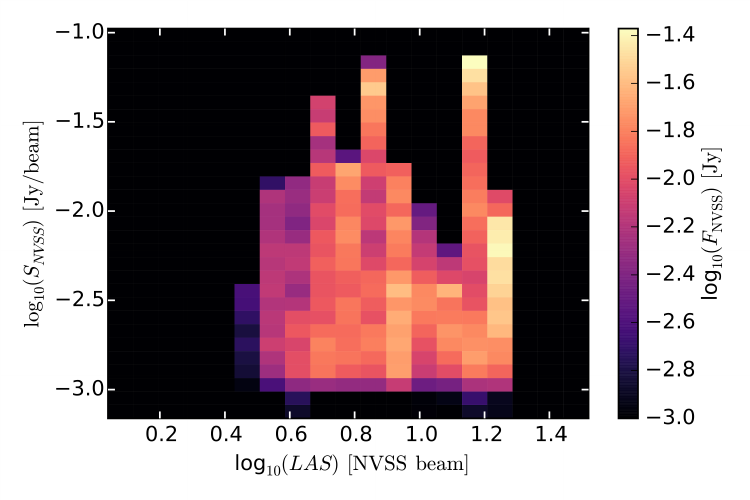}    
    \caption{Binned relic surface brightness  over the projected distance to the cluster center 
    $D_\mathrm{proj}$ (left-hand panel) and LAS\,[NVSS $\mathrm{beam}$], respectively 
    (right-hand panel). The color scale indicates the logarithmic flux per bin.}
    \label{Fig:Histo}
  \end{figure}

   The flux density, smoothed by the NVSS beam, is a direct measure of the relic flux on the sky.
   Fig.~\ref{Fig:Histo} shows that relics have typical surface brightness, size and projected distances 
   to the cluster center, which are correlated.
   We recover the steep drop of of the relic flux contribution towards 1.3 times the detection threshold.
   
   Additionally, we want to point out two interesting details: Firstly, most of the radio emission resides in structures with fluxes only 1-5 times the relic detection threshold at projected distances of about 1.2\,Mpc.
   This leads to  the question whether more relic flux resides in yet unidentified low surface brightness structures. Secondly, the relic the surface brightness seem to be attenuated towards larger projected distances and smaller angular sizes,
   meaning that bright radio relics should not appear at arbitrarily large distances.

\section{Acknowledgement}
JG and MH acknowledge support by the Deutsche Forschungsgemeinschaft within the Research Group \textit{Cosmic Magnetism}.
SEN also thanks the Deutsche Forschungsgemeinschaft for support under the grant NU 332/2-1.

\end{document}